\documentstyle[preprint,aps]{revtex}
\tightenlines

\begin{document}
\title{{\large {\bf Metal-insulator transition in 2D: a role of
the upper Hubbard band}}}
\author{{V.I.Kozub and N. V.
Agrinskaya}}
\address{A. F. Ioffe Institute,
 194021 Saint Petersburg, Russia }

\maketitle
\date{\null}

\begin{abstract}
\baselineskip=3.5ex

To explain the main features of the metal-insulator 
transition (MIT) in 2D we suggest a simple model taking into account 
strongly localized states in the band tail of 2D conductivity band with a 
specific emphasize of a role of doubly-occupied states (upper Hubbard
band (UHB)). The metallic behavior of resistance is explained as result of 
activation of localized electrons to conductance band leading to a 
suppression of non-linear screening of the disorder potential. The 
magnetoresistance (MR) in the critical region is related to
depopulation of double occupied localized states also leading to
partial suppression of the nonlinear screening. The most informative
data are related to nearly activated temperature dependence of MR in
strongly insulating limit (which can be in particular reached from
the metallic state in high enough fields). According to our model
this behavior originates due to a lowering of a position of chemical 
potential in the UHB due to Zeeman splitting. We
compare the theoretical predictions to the existing experimental data
and demonstrate that the model explains such features of the 2D MIT
as scaling behavior in the critical region, saturation of
MR and $H/T$ scaling of MR in the
insulating limit. The quantitative analysis of MR in 
strongly insulating limit based on the model suggested leads to the
values of $g$-factors being in good agreement with known values for 
localized states in corresponding materials.

\end{abstract}

\textwidth=6.0in

%\narrowtext

\section{Introduction}
The problem of apparent metal-insulator transition in 2D
structures observed for
different 2D systems including in particular
Si MOSFETs (see e.g.
\cite{Krav}, \cite{Sar}) and GaAs-AlGaAs
heterostructures (see e.g.\cite{heter},\cite{hole},
\cite{Yoon});
the extent review of the situation is
given in \cite{review})
is still far from complete understanding.
At the present time the situation is disputed between the two
possible scenarios. According to the first one
the unusual
metallic behavior which apparently contradicts the
scaling theory of localization \cite{scal} is
a real manifestation of some new physical mechanism.
According to the second one,
the metallic conductivity is still expected to be suppressed at low
enough temperature while the behavior observed until now is related
to the mechanisms of more conventional nature 
like temperature-dependent disorder
\cite{Altshuler} or magnetic-field driven disorder \cite{Pudalov2}.
Despite of various mechanisms proposed, according to the conclusion of
a review paper \cite{review} "while each of these is capable
of explaining one  or
another part of experimental observations, none of them provide a
comprehensive picture".

The theoretical approaches
developed until now were mostly concentrated on the metallic
state and discussed an effect of disorder and electron-electron
interactions on this state while to the best of our knowledge
practically no attempts were made to
look at the problem starting from the insulating limit. 
The role of localized states
was to some extent discussed in the papers \cite{Altshuler},
\cite{Pudalov2}, \cite{Pudalov1}, \cite{Pudalov3}
where an effect of traps in oxide layer or
interface region of Si MOSFETs was considered.
However these localized states were actually external with respect to
the states of 2D conduction band and only played a role of additional
scattering centers while we mean
the states responsible for
the 2D transport deep in the insulating limit.  
Then, the traps - although by no doubt
important for Si MOSFETs - can be hardly considered for other
systems exhibiting the 2D MIT behavior.

To our opinion, the clue to the problem of the 2D MIT can be related to
the MR in strongly insulating limit
(which can be in particular reached from initially
metallic states in high enough fields). As it is known,
the strong positive MR  in parallel
magnetic fields leading to a suppression of metallic
state in the "critical" region stays to be one the puzzles of the 2D
MIT. We especially emphasize that strong positive MR
persists deep in the insulating state and is exhibited by the systems
which were perfectly insulating at $H = 0$ \cite{review}.
The specific feature of MR in this regime is
the fact that its temperature dependence is close to the
Arrenius law even if
for $H = 0$ the system has exhibited variable range hopping of
Efros-Shklovskii type (which was observed for Si MOSFETS
\cite{Klapwijk1} and $n$-type GaAs
heterostructures \cite{Issai}). We believe that the
renormalization approaches starting from the metallic state
(see e.g. \cite{Fin}),
possibly explaining
MIT and suppression of metallic phase by magnetic field, fail to
explain this MR deep in insulating phase.

As it was shown earlier \cite{ours}, the activated behavior of
MR in hopping regime is a signature of the
hopping over the states of the UHB, that is over the
doubly-occupied localized states. Because of the on-site spin
correlations the Hubbard energy acquires in this case the Zeeman
addition $g\mu_B H$ ($g$ being the $g-$factor) and thus
magnetic-field dependent addition is a universal one (at least for
$g \mu_B H >> T$). Basing on these facts,  we suggested that
the UHB plays an important role in the problem of 2D
MIT \cite{condmat}. Note that a possible role of
the multiply-occupied electron states in 2D MIT
was also discussed in \cite{Klap}.

In what follows we suggest a model which starts from the localized
limit of the 2D MIT rather than from the metallic one
and thus is concentrated mainly on the "insulating" and "critical" regimes.
We believe that in the systems exhibiting the 2D MIT
there exists a tail of strongly localized states below the bottom
of 2D conductance band with localization lengths $a$
smaller than predicted by scaling theory of localization
\cite{scal} starting from the metallic regime; 
localization length $a$ as a function of energy for these
deep states is expected to have a critical divergency with energy increase
with a cut-off imposed by STL at some threshold energy. 
The highest of these states
according to the predictions by Kamimura \cite{Kamimur} 
are the double-occupied
ones. Namely, the energy dependence of the Hubbard energy
leads to formation of the peak of the UHB
close to the bottom of the conductance band \cite{Kamimur}.
The metallic behavior of resistance in our model
is related (like in the model
\cite{Altshuler}) to temperature-dependent disorder
originated in our case
due to activation of electrons from the strongly localized states to the
conductance band which leads to 
partial undressing of the disorder
potential. The MR in the metallic regime
is expected to be due to redistribution of electrons between
different doubly- and single occupied states (driven by the
Zeeman addition to the energy of
double-occupied states) which also leads to partial suppression of
screening of the disorder potential. (The similar 
concepts were 
suggested in  Refs.\cite{Pudalov3}, \cite{Pudalov2} where 
depopulation of filled traps to 
delocalized states due to
an increase of energy of the traps in magnetic field
was considered). 
This simple model is shown
to explain - at least qualitatively - most of the
features of the observed 2D MIT (including scaling of resistance in
critical region, $H/T$ scaling of MR in the insulating
limit and saturation of MR at large $H$).
Basing on the fact that in the strongly insulating limit the values
of $g$-factor are not expected to be renormalized by Fermi-liquid
effects,  we have compared
some of existing MR data with predictions of our
model. A good agreement between g-factors extracted according
to our predictions and their known values
gives to our opinion a
strong support to our model.

\section{Formulation of the model}

We {\it assume} that the 2D conductance band has a
band tail of strongly localized states originating due to
localization of carriers in the potential relief
imposed by the disorder potential. We
believe that the
localization length for these states behaves as
\begin{equation}\label{a}
 a \propto (\varepsilon_m - \varepsilon)^{-\nu}
\end{equation}
where $\varepsilon_m$ is some energy which for 3D would
correspond to mobility edge while $\nu$
is an index of localization length.

Actually STL
predicts for 2D an absence of any mobility edge and allows only
some threshold values of conductance 
separating regimes of weak and strong localization.
The localization length is expected to be exponentially
dependent on the value of conductance in the weakly localized
regime and to decrease when the conductance tends to the
critical value mentioned above. Thus one may expect that
for a given degree of disorder there is some minimal value of
the localization length
${\tilde a}$ available in the STL (of the order of the mean free
path).

In our model we will assume that the localization lengths of the
strongly localized states described by Eq.\ref{a} is
{\it smaller} than $\tilde a$. This assumption corresponds to a
cut-off of the divergency in r.h.s. of Eq.\ref{a} at some
$$ \varepsilon = {\tilde \varepsilon_m} < \varepsilon_m $$
while for higher energies the situation can be controlled by the
STL. 
In other words, at $\tilde \varepsilon_m$ we have a crossover
from strongly localized states
described by Eq.\ref{a}, to the states described by STL.
While the states in the vicinity of $\tilde \varepsilon_m $
are still strongly localized, the localization length exponentially
increases with an increase of $\varepsilon - {\varepsilon_m}$.
 
In what follows we will also assume that the difference
$\varepsilon_m - {\tilde \varepsilon_m} $ is small with respect to
characteristic energy scales and can be neglected. 
It means that the width of the band of states strongly localized in accordance
to STL is much smaller than the width of the band of strongly
localized states described by Eq.\ref{a}.
Note that the behavior of this sort is expected just for the
"new" 2D structures exhibiting MIT since these structures
exhibit high mobilities and, correspondingly, 
the values of $\tilde a$ for them are expected to be large.
Having these considerations in mind in what follows
we will still  
use the term "mobility edge" for the energy $\varepsilon_m$,

As it is known, the systems exhibiting 2D MIT are characterized
by strong electron-electron interactions.
In our model
we will directly take into account the intrastate
electron-electron interactions which were shown in \cite{Kamimur}
to be most important. Following the concepts by
Kamimura \cite{Kamimur}, we take into account the
energy dependence of the Hubbard energy $U$:
\begin{equation}
U \simeq \frac{ e^2}{\kappa a}
\end{equation}
where $a$ is energy-dependent localization length while
$\kappa$ is the dielectric constant.
While for 3D
it is known to be strongly dependent on
localization length (through the wave-vector dependence)
there are convincing arguments to assume that
for 2D the value of $\kappa$ independent or almost independent
of $a$ and close to its bulk value; these arguments were given
in the paper \cite{Shklov} where the exponent in the relation
$\kappa \propto a^{\alpha_2}$ was estimated as $\alpha_2 < 0.2$
However we still consider a possibility of $\alpha_2 \neq 0$
and thus we have
\begin{equation}
U \propto (\varepsilon_m - \varepsilon)^{u}
\end{equation}
where $u > \nu $.

As it is shown in the Appendix 1, the energy dependence of the Hubbard
energy has significant consequences for the form of the upper
Hubbard band. Namely, for strong electron-electron repulsion
this band is peaked near the energy $\varepsilon_m$ \cite{Kamimur}
and the peak is characterized by a steep decrease of DOS for
$\varepsilon < \varepsilon_{-,c}$ where $\varepsilon_{-,c}$
is some threshold energy  (see Fig. 1).
When $\mu =\varepsilon_m$, that
is  at the point of metal-insulator transition
(in the sense of strong localization) all states below the
mobility edge (possibly except of the states deep
in the band tail) are doubly-occupied.
This fact and a presence of a peaked DOS of $D^-$ band
below $\varepsilon_m$ allow us to conclude that the
metal-insulator transition takes place in the $D^-$ band
and its presence significantly affects the features of the transition.

We would like to note that the $D^-$ states in our case,
strictly speaking, are different from "standard" impurity $D^-$
states similar to negative hydrogen ion. In the latter case the
two electrons are in the field of a single positive elementary
charge. For the "tail states" it is not the case since the
electrons are considered to be localized in the potential wells
created by charge density outside of the 2D sheet like the
charges of the interface traps in the case of silicon MOSFETs or
spatial charge fluctuation as in the case of GaAS-AlGaAs
heterostructures.
In this case the localization centers are by no means of the
Coulomb nature. Thus the double occupation of some localized
state does not mean a creation of negatively charged Coulomb
center as well as single-occupied state is not similar to
neutral center. (Nevertheless for the simplicity we will
still use the notations $D^0$ and $D^-$ for singly- and
doubly-occupied states.)
As it was mentioned above, the occupation
of the localized states by single electrons or by electron pairs
for the external "probe" particle is reduced to the effect
of non-linear screening of the initial disorder potential.

As for the rest of electron-electron interactions,
we believe that the latter can be taken into account
as
a nonlinear screening of the initial disorder
potential by electrons trapped to these states with an account of
electron-electron correlations.
As a result, the form of the band tail and
the position of the mobility edge are expected to depend on the
electron concentration $n$.
For small variations of electron density $\delta n$ linearizing the
dependence in question one obtains
\begin{equation}
\delta \varepsilon_m = \gamma \delta n
\end{equation}
We believe that the parameter $\gamma$ is scaled with the magnitude of
the disorder potential subjected to the nonlinear screening and,
correspondingly
can be estimated
on the base of estimates of efficiency of nonlinear screening as
$$ \gamma \sim {e^2\over \kappa n^{1/2}} $$
Thus both the chemical potential and
the mobility edge are expected to be
sensitive to $n$ and one has
\begin{equation}\label{evolut}
\delta (\varepsilon_m - \mu) = (\gamma + g_t^{-1})\delta n
\end{equation}
where $g_t$ is some total density of states at $\varepsilon =
\varepsilon_m$.
Thus the critical
value
of $n = n_c$ corresponds to $\mu(n) = \varepsilon_m(n)$.

We assume that the initial ("bare")  disorder energy is comparable to
electron-electron interaction energy $e^2n_c^{1/2}/\kappa$
which ensures effective
non-linear screening. This factor arises in "new" systems with
larger mobilities and weaker disorder.
In  systems with larger degree of disorder the
latter - at least in the insulating limit -
is less sensitive to electron occupation.
Another feature of these systems is a larger spatial scale of the
disorder potential which makes even bare localization length
to be large enough.

While the occupation of the localized states corresponds to
non-linear screening of the disorder potential, the occupation
of the states above $\varepsilon_m$ drives the system into
regime of linear screening according to the considerations given
in \cite{Efros}. Thus the metal-insulator transition is
accompanied by a transition from nonlinear screening to linear
one; however even in the metallic state - at least in the
"critical" region - one expects that short-range harmonics of
the disorder potential are still screened by localized electrons.

It is interesting to note that in the critical region when
$g_t$ can be considered to be equal to the 2D band DOS $=
2m/\pi\hbar^2$, the ratio of the terms $\gamma$ and $g_t^{-1}$
in the r.h.s. of Eq.\ref{evolut} is of the order of the
interaction parameter $r_s = 2^{1/2}me^2/\kappa \hbar p_F$
($p_F $ being the Fermi momentum) and thus the metal-insulator
transition is dominated by the shift of the mobility edge.

\section{Temperature behavior of resistance}

In our model we relate the temperature behavior of resistance in
metallic state, in accordance to \cite{Altshuler}, to
temperature dependent disorder.
Namely, we assume that the ionization of strongly localized
states below $\varepsilon_m$ leads to additional scattering for
the mobile electrons qualitatively in the same way as ionization
of occupied donor state create an additional charged scatterer.
Indeed, the nonlinear screening of the disorder potential by the
localized states is reduced when the number of localized
electrons is reduced. We appreciate that quantitatively
the contribution of the states with large localization lengths
is not expected to be similar to contribution of
donor state efficiently separated from the other localized
states, but believe that the qualitative similarity still exists.
However the efficiency of the scattering  by ionized
localized state is expected to
depend on the localization length of the state and in general
to increase with decrease of $a$ because of the larger momentum
transfer.

Thus the total "classical" resistance can be written as
\begin{equation}
\rho = \rho_0 + \rho_1
\end{equation}
where
\begin{equation}
\rho_1 = {\hat \rho_1} ({T \over {\tilde T}})^q\exp( - ({T_{s}
\over T})
\end{equation}
Here the exponential describes activation and thus according to
our considerations $T_s \sim \mu - \varepsilon_m$
while the $T$-dependence of  preexponential
(chosen as some power function)  is related to dependence of
scattering efficiency on $a$ and thus on the energy of the
state.
As it is seen, this expression  coincides with one
given in \cite{Altshuler} for the traps below the Fermi level.

Note that in principle the "residual" resistivity $\rho_0$
is also temperature dependent due to activation-induced increase of
concentration of mobile carriers. To take this dependence into
account we apply the Drude-like relations
\begin{equation}
\rho_0 = \frac{1}{n_m\mu_0},\hskip1cm \rho_1 = \frac{1}{n_m\mu_1}
\end{equation}
where $1/\mu_0$ and $1/\mu_1$ are the inverse mobilities controlled
by the "residual disorder" and "temperature-dependent disorder",
respectively. With a simplest assumption $\partial (\mu_1^{-1})/
\partial n \equiv {\tilde \mu}^{-1} = const$ (that is
with a neglect of a dependence of scattering efficiency on $a$)
one easily obtains
\begin{equation}\label{TCR}
\rho(T) - \rho(0) = \left( \frac{1}{n_{m0}} -
\frac{1}{n_m(T)}\right) \left(\frac{n_{m0}}{\tilde \mu} - \frac{1}{\mu_0}
\right)
\end{equation}
where $n_{m0} = n_m(T = 0)$.

\section{Magnetoresistance}

The simplest situation corresponds to dielectric limit
($\mu < \varepsilon_m$) with an additional assumption
$g_0(\mu)  > g_{-}(\mu)$.
In this case  (see Fig.1,b and also Appendix 2)
the main effect
of the magnetic field is related to
the shift of $\mu$ in the $D^-$ band with magnetic field increase.
For $T = 0$ this shift is equal
to $\delta \mu(H) = \mu_B g H$ (where the value of $g$
corresponds to localized states) while for finite temperatures
the position of the chemical potential is described as (see
\cite{ours})
\begin{equation}\label{HTscal}
\frac{\delta
\mu (H)}{T}=\ln (\cosh \frac{\mu _{0}gH}{2T})
\end{equation}
In the regime considered
the dominant contribution
to conductivity is related to hops within the $D^-$ band
from doubly occupied states with $\varepsilon = \mu$
since the hops
with spin flips from the single occupied states to single occupied
states are not allowed (as in the case of mechanism considered by
Kamimura \cite{Kamimur}).

We will begin with the case when there is
no Coulomb gap at the Fermi
level that is when the interstate
Coulomb interactions are neglected.
Note that for the hopping in the band tail with small DOS
the situation  crucially depends on the relation
between the width of effective hopping band
$\delta \varepsilon$
\begin{equation}
\delta \varepsilon \simeq \frac{T^{2/3}}{(g_c(a/2)^2 )^{1/3}}
\exp  \frac{(\varepsilon_c - \mu)^2}{3\varepsilon_1^2}
\end{equation}
estimated for a standard
VRH way for the value of DOS equal to $g= const = g(\mu)$
(let us restrict ourselves by the VRH of the Mott type) and
and the energy scale $\varepsilon_1$ characterizing energy dependence
of DOS near the Fermi level. If $\delta \varepsilon < \varepsilon_1$
then one deals with a standard Mott-type hopping estimated for the
corresponding density of states being approximately symmetric around the
chemical potential.

However when the opposite unequality holds which can be rewritten as
\begin{equation}\label{activne}
\frac{(\varepsilon_c - \mu)^2}{\varepsilon_1^2} >
2 \log \left((g_c\varepsilon_1)^{1/2}\frac{a}{2}\frac{\varepsilon_1}
{T}\right)
\end{equation}
the hopping is expected to be dominated by electron hops from the Fermi
level to empty states high above the Fermi level where the density
of states drastically increases.

We believe that it is legitimate to consider this sort of hopping
in a standard for the VRH way, that is optimizing the hopping probability
with respect to the typical hopping distance $\sim N_{\varepsilon}^{-1/2}$
(where $N_{\varepsilon}$ is a density of sites with
energies less than $\varepsilon$) and to the activation energy
$\varepsilon - \mu$. Here we consider a case of
the Gaussian tail while the generalization for the exponential tail
is straightforward.
Taking into account that
$$ N_{\varepsilon} = \int_{\mu}^{\varepsilon} {\rm d}\varepsilon'
g_c \exp - \frac{(\varepsilon_c - \varepsilon')^2}{\varepsilon_1^2}
\sim g_c \varepsilon_1
 \exp - \frac{(\varepsilon_c - \varepsilon)^2}{\varepsilon_1^2} $$
(where the integral is considered to be dominated
by the upper limit)
we obtain the following condition for the activation energy
$\varepsilon^{*}$:
\begin{equation}
\varepsilon^{*} \simeq \varepsilon_c - \frac{\varepsilon_1}{\sqrt 2}
\cdot
\lbrack \log \left( \frac{\varepsilon_1}{T}(g_c\varepsilon_1)^{1/2}
\frac{4 a}{2}\right) + \log \frac{\varepsilon_1}{\varepsilon_c -
\varepsilon^{*}}
\rbrack^{1/2}
\end{equation}
Since, at it is seen, $\varepsilon_1/(\varepsilon_c - \varepsilon^{*})
\simeq 1$ we will neglect the last term in $\lbrack \rbrack$.
Correspondingly, the temperature behavior of conductivity
is expected to obey the law
\begin{equation}\label{activ}
\sigma \propto \exp - \frac{
 \varepsilon_c - \mu - \frac{\varepsilon_1}{\sqrt 2}
\lbrack \log \left( \frac{\varepsilon_1}{T}(g_c\varepsilon_1)^{1/2}
\frac{4 a}{2}\right)
\rbrack^{1/2} }{T}
\end{equation}
As it is seen, the activation energy logarithmically depends on
temperature and slightly decreases with temperature decrease.
The difference with respect to a pure Arrenius law originates
due to a finite density of states in the tail.

If one includes the Coulomb gap into considerations, it would lead to a
quadratic gap near the Fermi level, but the strong energy dependence of
"bare" density of states does not allow for this gap to develop up to its
nominal value according to the value of $n(\mu)$ due to a cut-off at
energies $\sim \varepsilon_1$ controlling the DOS decay.

Now let us turn to the magnetic field dependence which is the most
important for us.
As it is clearly seen, for strong magnetic fields the position of the
Fermi level is significantly lowered which leads to
a dramatic
decrease of DOS at Fermi level.
In particular, the following scenario is possible. If
at $H=0$ the DOS at the Fermi level is large enough 
(see Fig.1,a) for the unequality
opposite to the one of Eq.\ref{activne} to obey,
the standard VRH
of Mott type or of Efros-Shklovskii type is observed.
In particular it is expected for the case when initial position
of the Fermi level corresponds to the peak region of $D^-$ band.
However
an application of magnetic field shift the position of $D^-$
band with respect to the Fermi level and finally
can lead to nearly activated behavior according to Eq.\ref{activ} when
$\mu_B g H > \varepsilon_1$.

One notes that this hopping via the states of $D^-$ band
actually competes with hopping over  (spin-polarized) $D^0$ band, that is with
hopping from occupied to empty states.
Correspondingly, when magnetic field-dependent contribution of
$D^-$ band becomes smaller than the contribution of hopping over
the $D^0$ states, MR saturates.

If we start from the metallic regime (see Fig.2,a)
the variation of $\varepsilon_m$ with magnetic
field is of principal importance.
Moreover, one expects the MR in this case to be mostly
related to the variation in question. Note that
such a behavior is in agreement with the ideas of "magnetic
field driven disorder" formulated in \cite{Pudalov2}.
An important factor is related to the fact that the metallic
state in principle corresponds to linear screening while the spin
polarization of the mobile electrons is not expected to produce
a strong effect on the disorder potential (see e.g.
\cite{dolgopolov}. However we believe that
at least in the critical region the linear screening still co-exist
with the nonlinear one and, in particular,
the localized $D^-$
states well below $\varepsilon_m$ with relatively small
localization lengths $a \leq n_c^{-1/2}$ can not be efficiently
screened by the mobile carriers.

Thus the devastation of the $D^-$
states
corresponds to
blocking of the nonlinear screening which is similar to creation
of new scattering centers. So the polarization of localized
states can lead to significant shifts of $\varepsilon_m$ (see Fig.2,b).
The more detailed arguments are given in the Appendix 2.

Estimating the variation of the  electron concentration
due to a suppression of the $D^-$ states with magnetic field increase
as
$$ \delta n \simeq g_{-} (\mu_B g \delta H)  $$
and applying the arguments similar to those leading to
Eq. \ref{evolut} one obtains an estimate for the shift
of the mobility edge as
\begin{equation}\label{evolutH}
\delta (\varepsilon_m - \mu) \simeq \gamma g_{-} \mu_B g \delta H
\end{equation}
So one concludes that an increase of magnetic field can lead to
a change of a sign of the l.h.s. of Eq.\ref{evolutH}
from negative to positive
which corresponds to a suppression of metallic state.

One concludes that in this regime the
saturation of the MR corresponds to stabilization
of $\varepsilon_m$ which corresponds to total polarization of the
electron system including the localized states.
The saturation field
$H_{sat}$ corresponds to the lifting of the bottom of
$D^-$ band to chemical potential.

As for the final state of the system for $H >
H_{sat}$, it obviously depends on the initial electron concentration.
If it is large enough, in the final state the chemical potential
is still larger than $\varepsilon_m$ and the system persists in the
metallic state. Moreover, we believe that deeply in the metallic
state the linear screening dominates and the MR can be
related to the mechanisms of a sort of considered in \cite{dolgopolov}.
Thus we do not intend to
analyze this regime in detail.

\section{Discussion}

Now let us compare predictions of our model with existing
experimental data.
We shall start with temperature behavior of resistance.
According to our Eq.\ref{TCR}
the sign of
temperature coefficient of resistance depends on the sign
of second bracket in Eq.\ref{TCR}.
Moreover, Eq.\ref{TCR} describes "metal-insulator transition"
within purely classical regime. Indeed, for
$$n_{m0} = {\tilde \mu}/\mu_0 \equiv n_{mc} $$
$\rho(T) = const$. For smaller values of $n_{m,0}$
the temperature behavior of resistance is dominated by
activation of electrons from localized states to delocalized ones
(in accordance to observations reported in \cite{Pudalov3}:
\begin{equation}
\rho \propto \exp (\frac{\varepsilon_m - \mu}{T})
\end{equation}
One notes that even for purely classical description our model
exhibits with exponential accuracy scaling of resistance
behavior $\rho(T, n_s)$ which is symmetric with respect to "critical" curve
with $\varepsilon_m - \mu = 0$ if one assumes a constant DOS
at the "critical region" since in the latter case (see Appendix 2)
$$ n_s - n_c = g_t (\mu - \varepsilon_m) $$
This behavior is in agreement with experiment (see e.g.
\cite{review}).
For strongly localized regime when $ \varepsilon_m - \mu$ is
large enough the system is in hopping regime. While deep in this
regime variable range hopping is expected which is in agreement
with experimental data (see e.g. \cite{review}).

Actually the quantum effects tend the system to insulator
for $T \rightarrow 0$ since the states above the cut-off energy
$\tilde \varepsilon_m$ discussed in the Section 2 are described by
STL. The corresponding scenario is
considered in detail
in
\cite{Altshuler} and thus the metal-insulator transition
discussed above is not a true one.

Now let us discuss the MR.
As we have mentioned above, we are mostly interested in the
insulating limit because, to our opinion, it can give more information
about the system properties than the metallic limit. Thus
we would like first to compare our predictions concerning the
MR in strongly insulating limit with existing
experimental data.

As it was reported (see e.g. \cite{review}, cite{Klapwijk1},
), the high-field
insulator regime typically exhibits nearly activated
temperature dependence of the magnetoconductivity.
We shall start from experiment reported in the paper
\cite{Issai}
where a delta-doped GaAs/AlGaAs heterostructure
was studied. Although strictly speaking this system
did not exhibited 2D MIT, it demonstrated strong positive
MR starting from the hopping regime.
A specific feature was a gradual transition from the variable range
Efros-Shklovskii hopping to nearly activated hopping.
In our previous paper \cite{condmat}
we have tried to fit the experimental curves according to the pure
Arrenius law ascribing the deviations from experimental behavior
to a neglect of the non-zero DOS in the band tail.

Now we are going to compare the experimental data of Ref.\cite{Issai}
with our present theory taking into account a Gaussian tail
of the $D^-$ band.
At the Fig. 3 we have plotted theoretical curves corresponding
to our Eq.\ref{activ} calculated for
$\varepsilon_c - \mu(H = 0) = 0.234 meV (H = 8T),
0,219 mev (H = 6T)$, $\varepsilon_1 = 0.14
meV$ , $g_0 a^2 = 0.93 meV^{-1}$, $g = 0.12$.
As it is seen, there is a good agreement between experimental
data and our theoretical prediction which supports our present
interpretation.

The value of the effective $g$-factor
is about 4 times lower than the
handbook values for GaAs. However, one has in mind that we deal with
AlGaAs-GaAs heterostructure rather than with the bulk GaAs. The
$g$-factor values for AlGaAs quantum wells were calculated theoretically
\cite{Ivchenko}. It was shown that due to the fact that in such
structures one has a mixture of GaAs states (for which the $g$-factor
$\sim -0.45$ is negative) and AlGaAs states (where $g$ -factor is
positive) the effective $g$-factor depends on the well width.
In the case of gated heterostructure
under discussion, the effective width of the potential well is
in particular controlled by the gate voltage $V_g$ and thus
the effective $g$-factor is expected to depend on $V_g$.
The corresponding behavior was considered in some detail in our
previous paper \cite{condmat} and was shown to be in agreement
with the experimental data of \cite{Issai}.

In addition, we have compared our results with MR
data by Simonian et al.\cite{Sar}. The data obtained for Si
MOSFETs demonstrated a suppression of metallic state in strong
parallel magnetic fields and transition to strongly localized
state. Since our model gives most definite quantitative predictions
for insulating limit, we analyzed the data on temperature
behavior of MR
reported in \cite{Sar}
for the strongest magnetic fields (1.4 and 1.2 T).
On Fig.4 we present both experimental data and theoretical
curves calculated on the base of Eq.\ref{activ} with the values
of the parameters $\varepsilon_c - \mu = 0.147 meV
(H = 1.4 T), 0.126 meV (H = 1.2 T) $
$\varepsilon_1 = 0.1 meV$ $g_0a^2 = 1.2 meV^{-1}$, $g = 1.7$. As it is seen,
there is a good agreement between theory and experiment. The
value of $g$ factor estimated in this way is close to
the textbook values for Si. Note that in the insulating limit
Fermi-liquid renormalization of $g$-factor is irrelevant.

Following the same ideas we have also analyzed the
MR in the insulating
limit for the p-type
GaAs-AlGaAs heterostructures exhibiting the metal-insulator
transition \cite{Yoon}. The extracted value of $g$-factor
has appeared to be equal to 0.025-0.04 which is in agreement
with the fact that the expected value of $g$ factor for heavy
holes is zero and the effect is related to admixture of light
hole states.

The paper \cite{Sar} reported a suppression of the metallic phase at
strong magnetic field followed by a saturation of resistance, the
behavior exhibited $H/T$ scaling while the most representative data
corresponded to the dielectric side of MIT.
The corresponding behavior is in agreement with our
Eq.\ref{HTscal} predicting the scaling for the MR
activation exponent.

An important feature of the MR in the systems
exhibiting 2 D MIT is its saturation at high fields.
According to our model, the nature of this effect can be
different for different regimes.

For the systems initially being in the strongly
insulating limit   (when the disorder potential is not
significantly affected by magnetic field), the saturation field
$H_{sat}$ is related to a competition between contributions of
$D^-$ and $D^0$ subbands and thus depends mostly on the forms
of the subbands.

For the systems initially being in the "critical" region
(of no matter on what side of the transition) the initial
evolution of the MR is related to shift of
$\varepsilon_m$, although at high fields a competition
between hopping processes in $D^-$ and $D^0$ bands is still possible.
In both regimes mentioned above the value of $H_{sat}$
is expected to be sample-dependent.
Note that such a behavior was reported in \cite{Pudalov3}.

Although from the very beginning we have not intended to discuss
properties of the systems where initial electron density was
significantly larger than the critical one, we would like to
give some comments for this regime as well. In particular,
even for the systems being initially deep in the metallic regime
but still containing a significant density of the localized states
the value $H_{sat}$ is not expected to be universal since the
form of $D^-$ band is obviously sample-dependent. Moreover, the
possible exchange of electrons between the conducting band and the
states spatially separated from the conducting paths
(discussed in Appendix 2)
does not allow to relate the total electron
concentration to position of the chemical potential in conducting
band. In addition, one also notes that the value of $g$-factor for
mobile electrons is renormalized due to exchange interactions and is
concentration-dependent; in any case its value is different from the
corresponding value for localized states.
Thus we believe that the saturation field itself
for the metallic regime does not give a
proper information concerning to say the value of $g$-factor.
This conclusion is especially important for the samples with
concentrations close to the critical one.

Note that if $H_{sat}$ corresponds to the crossover from "metallic"
to "insulating" state then - at least for magnetic fields close to
$H_{sat}$ - the chemical potential crosses the peak of the $D^-$
band. This factor can explain the enhanced DOS
at the Fermi level for the "critical" state (with respect to the
standard one for the ideal 2D conduction band) observed in
\cite{Kravchenko}.

We would like also to note that the "final" insulating state
achieved in strong magnetic fields is different from the insulating
state existing for $H=0$ at small enough electron concentrations;
in particular, the latter situation is characterized by a presence
of a peak of $D^-$ band close to $\varepsilon_m$
which is absent for the former one. Correspondingly,
the electrons in metallic state close to the "critical" state
for $H=0$ are expected to suffer additional mechanism of inelastic
scattering related to activation of $D^-$ states which is absent for
spin-polarized situation. This latter factor can
enhance a role of weak localization for $H > H_{sat}$.
In any case, according to convincing calculations of
\cite{Altshuler2}, the behavior of the "critical" states
is strongly sensitive to the system parameters which -
with an account of the factor mentioned above - explains
a difference in behavior of the "critical states" for the cases of
$H = 0$ and $H > H_{sat}$ reported in \cite{Kravchenko}.

We believe that the surprising evolution
of temperature coefficient of resistance in course of magnetic field
increase ("metallic-like" in $H= 0$ and $H > H_{sat}$ and
"insulating-like" for some intermediate fields close to $H_{sat}$)
reported in \cite{Klapwijk1} can be related
to evolution of a position of $D^-$ band with respect to the
Fermi level. Indeed, one can expect that the presence of
quasi-localized states at the Fermi level at the moment when the
peak of the $D^-$ band coincides with $\mu$ can significantly
affect both classical and quantum contribution to resistance.

\section{Summary}
We have demonstrated that the main features of the 2D MIT
apparently observed recently in different systems can be
explained within the framework of
the simple two-band model. The latter is based on
the assumption that the transition-like behavior takes
place in the disorder-induced "tail" of the 2D conductance band
where the lower, strongly-localized states are not described by
scaling theory of localization starting from the large conductance limit. 
An important role among these states is played by doubly-occupied
states coexisting in general with the single-occupied ones.
The role of the $D^-$ states is emphasized by the fact that
as a result of energy dependence of the Hubbard energy
their DOS is peaked near the bottom of the 2D conductance band.
We also emphasize a role of non-linear screening of the disorder
potential by localized electrons.
The "metallic-like" temperature behavior of conductivity
is related to activation of the localized carriers to
the conductance band leading to undressing of the disorder
potential due to the partial suppression of the non-linear screening.
The strong positive MR on the dielectric side
is related to a suppression of activated hopping to the peak
of $D^-$ band as a result of lowering of chemical potential
with respect to $D^-$ band. The suppression of the metallic
state in strong magnetic field is explained as a result of a
suppression of doubly-occupied localized states
participating in the non-linear screening
of the disorder potential. The comparison of the theoretical predictions with
existing experimental data exhibits at least qualitative
agreement while the values of $g$-factors extracted from the
MR data for strongly insulating limit according
to our model are in quantitative agreement with the values
known for the localized states in corresponding materials.

\section{Acknowledgements}

We are grateful to D.Khmelnitskii and I.Shlimak
for discussions and valuable
remarks and to V.M.Pudalov for sending us the manuscript of the
paper \cite{Pudalov2} before its publication.
The paper was financially supported by RFFI,
Grant N 00-02-16992.

\section{Appendix 1.
Form of the D- band}

The energy of the doubly-occupied state $\varepsilon_{-}$
is obviously given as
\begin{equation}\label{epsilon-}
\varepsilon_{-} = \varepsilon_0 + U_0(
\frac{\varepsilon_m - \varepsilon_0}{\varepsilon_m})^{u}
\end{equation}
Correspondingly, the DOS in the $D^-$ band is given as \cite{Kamimur}
\begin{equation}\label{g-}
g_{-}(\varepsilon_{-}) =
g (\varepsilon_0(\varepsilon_{-}))\left( \frac{\partial
\varepsilon_{-}}{\partial \varepsilon_0}\right)^{-1}
\end{equation}
where $\varepsilon_0$ is the energy of single-occupied state while
the functional dependence of $\varepsilon_0(\varepsilon_{-}$
is given by a solution of Eq.\ref{epsilon-}.

As it is seen, $g_{-}$ diverges at some $\varepsilon_{0,c}$
corresponding to vanishing of the derivative in Eq.\ref{g-}:
\begin{equation}
\varepsilon_m = u U_0 (
\frac{\varepsilon_m - \varepsilon_{0,c}}{\varepsilon_m})^{u - 1}
\end{equation}
This is related to the fact that the function
$\varepsilon_0(\varepsilon_{-})$ is a double-valued one and
vanishing of the derivative in question corresponds to a
coincidence of these solutions. One branch of the solutions
corresponds to the single-occupied states above
$\varepsilon_{0,c}$ while another one - to the states below
$\varepsilon_{0,c}$.

Then, one concludes that if
$$\varepsilon_m - \varepsilon_{0,c} \simeq
\varepsilon_m \left(\frac{u \varepsilon_m}{U_0}
\right)^{1/(u - 1)} $$
is small enough with respect
to the depth of the band tail, the lowest localized states allow
only single-electron occupation because the Hubbard energy for
these states appear to be too high.

The critical value of the energy of double-occupied state
$\varepsilon_{-,c}$ is related to $\varepsilon_{0,c}$
by Eq.\ref{epsilon-}.
To estimate the energy dependence of $g_{-}$ near the critical
value of $\varepsilon_{-}$ we take into account that for
$\varepsilon_0 $ close to $\varepsilon_{0,c}$
\begin{equation}
\frac{\partial \varepsilon_{-}}{\partial \varepsilon_0}
\simeq \frac{\partial^2 \varepsilon_{-}}{\partial
\varepsilon_0^2} (\varepsilon_0 - \varepsilon_{0,c})
\end{equation}
To rewrite this equation in terms of $\varepsilon_{-}$
we take into account that
\begin{equation}
\varepsilon_{-} - \varepsilon_{-,c} \simeq
\frac {\partial \varepsilon_{-}}{\partial
\varepsilon_0}|_{\varepsilon_0 = \varepsilon_c} + \frac{1}{2}
\frac{\partial^2 \varepsilon_{-}}{\partial \varepsilon_0^2}
(\varepsilon_0 - \varepsilon_{0,c})^2
\end{equation}
Having in mind that the first derivative vanishes, we are left
with
\begin{equation}
(\varepsilon_0 - \varepsilon_{0,c})
\simeq \left( \frac{1}{2}\frac{\partial^2
\varepsilon_{-}}{\partial \varepsilon_0^2}\right)^{-1/2}
(\varepsilon_{-} - \varepsilon_{-,c})^{1/2}
\end{equation}
where
$$ \frac{\partial^2 \varepsilon_{-}}{\partial \varepsilon_0^2}
\simeq \frac{u - 1}{\varepsilon_m} \left(\frac{u
U_0}{\varepsilon_m} \right)^{1/(u - 1)} $$
Correspondingly, we have for $\varepsilon_{-}$ close to
$\varepsilon_{-,c}$:
\begin{equation}\label{extr}
g_{-} = { 2^{1/2} g (\varepsilon_0(\varepsilon_{-})) \over
(\varepsilon_{-} -
\varepsilon_{-,c})^{1/2}}(\frac{\varepsilon_m}{u -1})^{1/2}(
\frac{\varepsilon_m}{u U_0})^{1/2(u - 1)}
\end{equation}
Note that in realistic situation the sharp edge of the DOS
singularity is expected to be smeared in some way.
In particular, the local fluctuations of the parameter $U_0$
are expected due to local fluctuations of the dielectric function.
It is natural to assume that the fluctuations of $U_0$ correspond
to Gaussian distribution with some variance $\delta U_0$.
As it can be easily followed, these fluctuations give rise
to Gaussian fluctuations of the energy $\varepsilon_{-,c}$
with the variance
$$\varepsilon_1 \sim \delta U_0 \frac{1}{u-1}
\left(\frac{u\varepsilon_m} {U_0}\right)^{1/u-1} $$
Averaging the form-factor function of
Eq.\ref{extr}
$$ \frac{1}{(\varepsilon_{-} - \varepsilon_{-,c})^{1/2}} $$
with respect to the fluctuations in question
one easily obtains that the averaged form-factor has a form
\begin{eqnarray}
\left((\varepsilon_{-} - \varepsilon_{-,c}) + \delta
\varepsilon_{-} \right)^{-1/2} \hskip0.5cm (\varepsilon_{-} >
\varepsilon_{-,c}
\nonumber\\
(\varepsilon_1
\exp -(\frac{\varepsilon_{-}-\varepsilon_{-,c}}{\delta
\varepsilon_{-}})^2) \hskip0.5cm \varepsilon_{-} < \varepsilon_{-,c}
\end{eqnarray}
As a result one expects the $D^-$ band to have its own tail
corresponding to non-zero but quickly decreasing DOS
at $\varepsilon_{-} < \varepsilon_{-,c}$:
\begin{equation}\label{d-tail}
g_{-} \simeq g_c \exp - (\frac{\varepsilon_{c} -
\varepsilon}{\varepsilon_1})^2
\end{equation}
where we have omitted indices $-$ at energy terms implying that
the function $g_{-}(\varepsilon)$
has energy argument corresponding to $D^-$ band.

\section{Appendix 2.
Occupation of $D^-$ band}

Now let us discuss the details of the occupation of $D^-$
band with an increase of the
concentration. If one neglects the states in the tail of $D^-$ band,
the occupation of this band starts when
the chemical potential touches its bottom, i.e. when $\mu =
\varepsilon_{-,c}$. The double-occupied states
first appear in the band of single-occupied states
near the energy $\varepsilon_{0,c}$ symmetrically with respect
to this value. Further increase of concentration leads
to rise of the position of chemical potential
in the $D^-$ band and to a broadening of the band of
double occupied states in $D^0$ band. Finally $\mu$ reaches
the value $\varepsilon_m$ and for $\mu \rightarrow
\varepsilon_m$ the gap between empty states and doubly-occupied
states in terms of energies $\varepsilon_0$ tends to zero.
In other words, at the moment of metal-insulator transition
(in the sense of strong localization) all states below the
mobility edge are doubly-occupied
(possibly except of the states situated deep enough).
Actually
one should have in mind that
according the considerations given above an increase of
concentration affects not only a position of $\mu$, but
a position of $\varepsilon_m$ as well and thus the picture
is more complex one than a simple occupation of
a band with a given density of states).

Formally at low temperatures one has for the chemical potential
the following equation:
\begin{equation}
n =  \int^{\mu} {\rm d} \varepsilon (g_0 + g_{-}),
\end{equation}
where $g_0$ correspond to DOS of single-occupied states in the
band tail, $g_{-}$ - to DOS of doubly occupied states discussed
above.
Since according to our
considerations mentioned above the very form of DOS in the
band tail in principle depends on $n$, the latter being given by
a solution of the general Hubbard Hamiltonian including both
intrastate and interstate interactions, the problem of
calculation of $\mu$ for a given $n$ is extremely difficult.

However for the small variations of $n$ near the critical
value $n_c$ one expects the linearization procedure
of the sort of Eq.\ref{evolut} is
possible.

Another problem of extreme importance for us is related to
variation of $\mu$ with a variation of magnetic field.
The presence of the field leads to the Zeeman addition $\mu_B g H$ in
the equation \ref{epsilon-}.
Since this addition is not energy-dependent, it does not
change a value of $\varepsilon_{0,c}$ discussed above, however
shifts the bottom of $D^-$ band $\varepsilon_{-,c}$
towards the mobility edge corresponding to majority spins.
Correspondingly, the number
of doubly-occupied states in $D^0$ band around the
energy $\varepsilon_{0,c}$ decreases. Finally the value
$\varepsilon_{-,c}$ coincides with $\varepsilon_m$
which obviously corresponds to
\begin{equation}\label{H-c}
 \varepsilon_m - \varepsilon_{-,c}(H = 0) = \mu_B g H
\end{equation}
After that moment only single-occupied states exist
below $\varepsilon_m$. Recall that the value of $\varepsilon_m$
is considered to be related to majority spins while for minorities
it is enhanced by the Zeeman splitting. Thus for the magnetic field
given by Eq.\ref{H-c} near the mobility edge for majority spins
the mobile electrons with majority spins co-exist at the same energy
with localized $D^-$ states which in particular gives additional
mechanism of resonant exchange scattering.

As for the position of the Fermi level, for a fixed number of
electrons it should be calculated with an account of a balance
between the number of electrons in single- and double occupied
states. For the system deep in the insulating state such analysis was
made in \cite{ours} with an assumption that the DOS of the double
occupied states at the Fermi level is much less than DOS for
single-occupied states. In this case the position of $\mu$ in the
$D^0$ band is fixed while its position in $D^-$ band is lowered
according to considerations given above.

The picture is not as simple if we start from metallic state.
If we would deal only with the band tail states, depopulation of
double occupied states (including the localized $D^-$ states)
would inevitably lead to rising of the chemical potential
for the majority electrons. Note that for conducting (delocalized)
electrons the value of $g$- factor can be renormalized due to
Fermi-liquid effects (see e.g. \cite{renormg}. )

Thus these simple considerations concerning the evolution of chemical
potential with increase of magnetic field
can not explain the strong positive
MR in the metallic state since the rising of chemical
potential for spin-polarized electrons is not expected to decrease
the conductivity significantly.
To our opinion, the MR can be related to evolution of
the position of the "mobility edge" $\varepsilon_m$ with increase of
field. This argument is supported by the estimate given in
Section 2 which shows that - at least at the
critical region - the evolution of $\varepsilon_m$
dominates the evolution of $\mu$ by a factor $r_s >> 1$.

Two possible scenarios can be discussed.
First, even if one restricts himself by the band tail states he can
expect that the devastation of the doubly-occupied  $D^-$ states
(that is redistribution of electrons between double - and single
occupied localized states)
lead to enhancement of the disorder potential and to increase of
$\varepsilon_m$. Indeed, one has in mind that occupation of the localized
states by electrons corresponds to screening of this potential
and the screening is expected to be more effective if there exists an
additional choice between double and single occupation (lowering the
total energy of the system). According to the arguments given
above the evolution of $\varepsilon_m$ dominates over the inevitable
increase of $\mu$ related to the devastation of $D^-$ states.

Another possibility corresponds
to a presence of the subband of
strongly localized states
assumed to coexist with the 2D band states (including the
band tail states), but being
spatially
separated from the band states. These localized states are
expected to have a mobility edge much larger higher than
$\varepsilon_m$ and much larger Hubbard energies preventing
the double occupation of these states. Such an assumption is
a realistic one since for small electron concentrations
the percolation character of electron transport is expected
due to spatial variation of $\varepsilon_m$ due to large-scale
potential \cite{Efros}.
In this case the electrons
from $D^-$ states are in particular redistributed to these states
decreasing the density of electrons in the band tail. Thus the
nonlinear screening of localized states becomes more effective
while the rising of chemical potential is suppressed.

As a result, the difference $\mu(H) - \varepsilon_m(H)$ can at some
$H = H_c$ change its sign which corresponds to a crossover to
insulating limit.

Although a possible shift in $\varepsilon_m$ with increase of carrier
concentration was disputed in \cite{Kravchenko} we would like to note
that the authors based their considerations on an absence of
significant shifts of $\varepsilon_m$ for small variations of
electron concentration around its critical value. To our opinion,
this argument does not hold for our picture since the
redistribution of electrons due to spin polarization is expected to be
pronounced in contrast to the case of small variation of $n_s$.

\vskip10cm
\vskip3cm {\large Figure captions}

\medskip

Fig.1,a  The structure of the subbands of
double-occupied (left) and single-occupied (right) localized states
for the electron occupation corresponding to insulating limit
in the absence of magnetic field. Hopping conductivity is
controlled by the states of $D^-$ band.

Fig.1,b  The occupation of the subbands depicted in Fig.1,a
for high magnetic field. Hopping is dominated by activation to
the peak of $D^-$ band
($\varepsilon_a = \varepsilon_c - \mu$)

Fig.2,a  The structure of the subbands of
double-occupied (left) and single-occupied (right) localized states
for the electron occupation corresponding to metallic state in
the "critical" region in the absence of magnetic field.

Fig.2,b  The structure of the subbands for high magnetic field
driving the system to strongly localized regime. Note that the
position of $\varepsilon_m$ in the $D^0$ band is considered to
be lifted with respect to the case $H = 0$ (moreover, the form
of the bands is also changed)
while the position of
$\mu$ in the $D^-$ band is lowered.

Fig.3   Resistivity versus temperature for Si $\delta $-doped GaAs-AlGaAs
heterostructure of Ref.\cite{Issai} for $B = $ 6 (full squares)
and 8 Tesla (crosses)
at $n=9.52\times 10^{10}$ cm$^{-2}$. Solid lines -
theoretical curves corresponding
to our Eq.\ref{activ} calculated for
$\varepsilon_c - \mu = 0.234 meV (H=8T), 0.219 meV (H=6T)$,
$\varepsilon_1 = 0.14
meV$ , $g_0 a^2 = 0.93 meV^{-1}$, $g = 0.12$.

Fig.4 Resistivity versus temperature for Si MOSFET of
Ref.\cite{Sar} for $B = $ 1.2 (full squares) and 1.4 (crosses)
at $n = 8.83 \times 10^{10} cm^{-2}$. Solid lines -
 theoretical
curves calculated on the base of Eq.\ref{activ} with the values
of the parameters $\varepsilon_c - \mu = 0.147 meV (H=1.4T), 0.126 meV
(H=1.2T)$.
$\varepsilon_1 = 0.1 meV$, $g_0 a^2 = 1.2 meV^{-1}$, $g = 1.7$.

\end{document}